\documentclass[10pt,letterpaper]{article}
\usepackage{opex3}

\begin{document}




\title{A superconducting focal plane array for ultraviolet, optical, and near-infrared astrophysics}

\author{Benjamin A. Mazin,$^{1*}$ Bruce Bumble,$^2$ Seth R. Meeker,$^1$ Kieran O'Brien,$^1$ Sean McHugh,$^1$ and Eric Langman$^1$}

\address{$^1$Department of Physics, University of California, Santa Barbara, California 93106, USA}
\address{$^2$NASA Jet Propulsion Laboratory, 4800 Oak Grove Drive, Pasadena, California 91109, USA}

\email{*bmazin@physics.ucsb.edu} 

\homepage{http://www.physics.ucsb.edu/~bmazin/} 

\begin{abstract} 
Microwave Kinetic Inductance Detectors, or MKIDs, have proven to be a powerful cryogenic detector technology due to their sensitivity and the ease with which they can be multiplexed into large arrays.  A MKID is an energy sensor based on a photon-variable superconducting inductance in a lithographed microresonator, and is capable of functioning as a photon detector across the electromagnetic spectrum as well as a particle detector.  Here we describe the first successful effort to create a photon-counting, energy-resolving ultraviolet, optical, and near infrared MKID focal plane array.  These new Optical Lumped Element (OLE) MKID arrays have significant advantages over semiconductor detectors like charge coupled devices (CCDs).  They can count individual photons with essentially no false counts and determine the energy and arrival time of every photon with good quantum efficiency.  Their physical pixel size and maximum count rate is well matched with large telescopes.  These capabilities enable powerful new astrophysical instruments usable from the ground and space.  MKIDs could eventually supplant semiconductor detectors for most astronomical instrumentation, and will be useful for other disciplines such as quantum optics and biological imaging.	
\end{abstract}



\section{Introduction}

Cryogenic detectors are currently the preferred technology for astronomical observations over most of the electromagnetic spectrum, notably in the far infrared through millimeter (0.1--3~mm)~\cite{Bintley:2010p4108,Niemack:2008p4173,Carlstrom:2011p4239}, X-ray~\cite{Kelley:2009p4337}, and gamma-ray~\cite{Doriese:2007p4284} wavelength ranges.  In the important ultraviolet, optical, and near infrared (0.1--5~$\mu$m) wavelength range a variety of detector technologies based on semiconductors, backed by large investment from both consumer and military customers, has resulted in detectors for astronomy with large formats, high quantum efficiency, and low readout noise.  However, these detectors are fundamentally limited by the band gap of the semiconductor (1.1 eV for silicon) and thermal noise sources from their high ($\sim$100~K) operating temperatures~\cite{Eisaman:2011p6692}.  Cryogenic detectors, with operating temperatures on the order of 100~mK, allow the use of superconductors with gap parameters over 1000 times lower than typical semiconductors.  This difference allows new capabilities.  A superconducting detector can count single photons with no false counts while determining the energy (to several percent or better) and arrival time (to a microsecond) of the photon.  It can also have much broader wavelength coverage since the photon energy is always much greater than the gap energy.  While a CCD is limited to about 0.3--1~$\mu$m, the new arrays described here are sensitive from 0.1~$\mu$m in the UV to greater than 5~$\mu$m in the mid-IR, enabling observations at infrared wavelengths vital to understanding the high redshift universe.  

This approach has been pursued in the past with two technologies, Superconducting Tunnel Junctions (STJs)~\cite{Martin:2006p4412,Hijmering:2008p4501} and Transition Edge Sensors (TESs)~\cite{Romani:2001p1716,Burney:2006p4521}.  While both of these technologies produced functional detectors, they are limited to single pixels or small arrays due to the lack of a credible strategy for wiring and multiplexing large numbers of detectors, although recently there have been proposals for larger TES multiplexers~\cite{Niemack:2010p4657}.  

Microwave Kinetic Inductance Detectors, or MKIDs\cite{Day03}, are an alternative cryogenic detector technology that has proven important for millimeter wave astrophysics\cite{Schlaerth:2010p3957,Roesch:2010p4655} due to their sensitivity and the ease with which they can be multiplexed into large arrays.  MKIDs use frequency domain multiplexing~\cite{Mazin:2006p5} that allows thousands of pixels to be read out over a single microwave cable. While the largest STJ array is 120 pixels~\cite{Verhoeve:2006p3383} and the largest optical TES array is 36 pixels~\cite{Burney:2006p4521}, the MKID arrays described below are 1024 pixels, with a clear path to Megapixel arrays.  The ability to easily reach large formats is the primary advantage of MKID arrays.  

In this paper we describe the first photon-counting, energy-resolving ultraviolet, optical, and near infrared MKID focal plane array.  These Optical Lumped Element (OLE) MKID arrays have significant advantages over semiconductor detectors like charge coupled devices (CCDs)~\cite{Smith:2011p4039}.  They can count individual photons with essentially no false counts and determine the energy and arrival time of every photon with good quantum efficiency.  Their physical pixel size and maximum count rate is well matched with large telescopes.  These capabilities enable powerful new astrophysical instruments usable from the ground and space.

\section{Detector design and fabrication}
\begin{figure}
\begin{center}
\includegraphics[width=1.0\columnwidth]{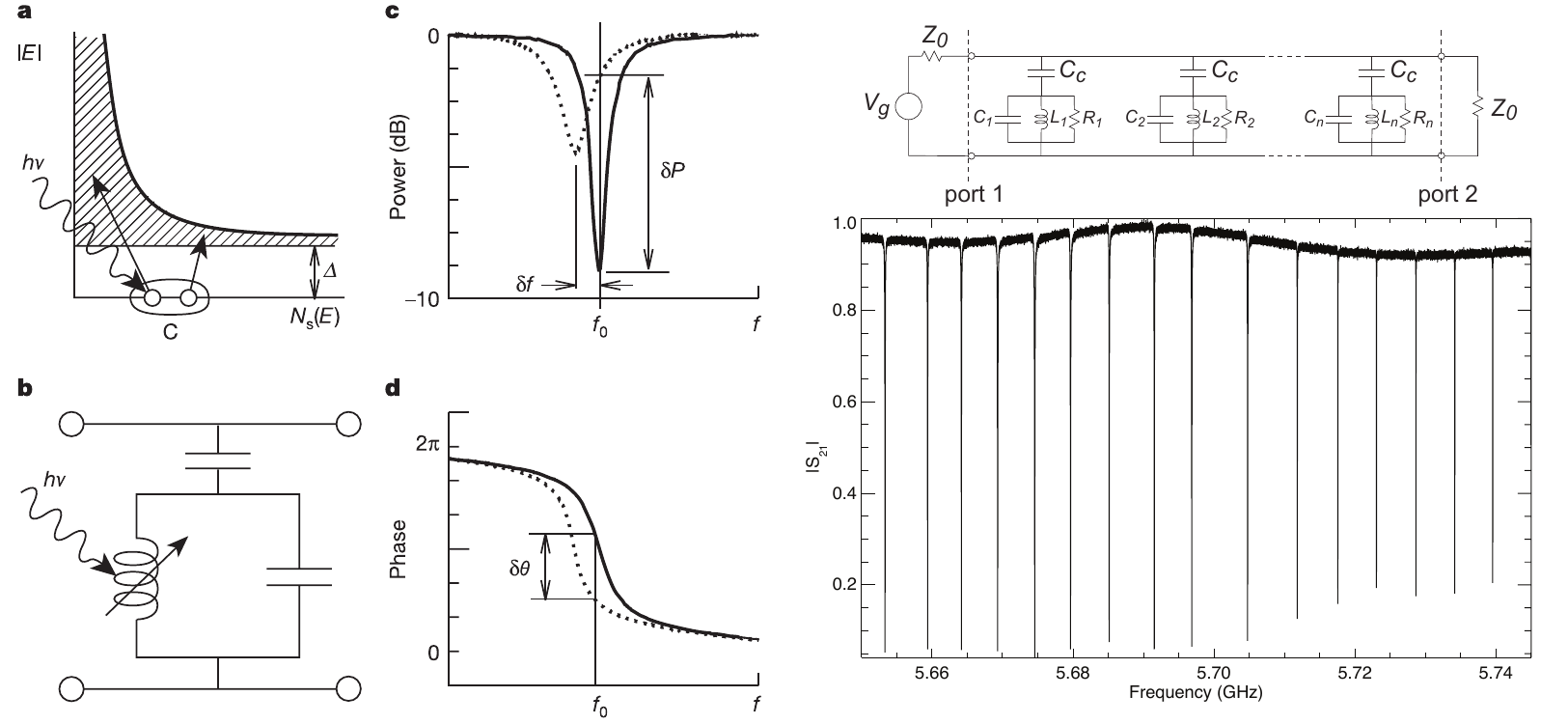}
\end{center}
\caption{Left: The basic operation of an MKID, from \cite{Day03}. (a) Photons with energy $h\nu$ are absorbed in a superconducting film, producing a number of excitations, called quasiparticles.  (b) To sensitively measure these quasiparticles, the film is placed in a high frequency planar resonant circuit.  The amplitude (c) and phase (d) of a microwave excitation signal sent through the resonator.  The change in the surface impedance of the film following a photon absorption event pushes the resonance to lower frequency and changes its amplitude.  If the detector (resonator) is excited with a constant on-resonance microwave signal, the energy of the absorbed photon can be determined by measuring the degree of phase and amplitude shift.  Right: The top panel shows the results the equivalent circuit of multiplexed MKIDs, and the bottom panel shows microwave transmission data from actual MKIDs with very accurate frequency spacing.} \label{fig:detcartoon}
\end{figure}

MKIDs work on the principle that incident photons change the surface impedance of a superconductor through the kinetic inductance effect~\cite{mattis58}.  The kinetic inductance effect occurs because energy can be stored in the supercurrent of a superconductor.  Reversing the direction of the supercurrent requires extracting the kinetic energy stored in the supercurrent, which yields an extra inductance.
This change can be accurately measured by placing this superconducting inductor in a lithographed resonator.  A microwave probe signal is tuned near the resonant frequency of the resonator, and any photons which are absorbed in the inductor will imprint their signature as changes in phase and amplitude of the probe signal.  Since the quality factor $Q$ of the resonators is high and their microwave transmission off resonance is nearly perfect, multiplexing can be accomplished by tuning each pixel to a different resonant frequency with lithography during device fabrication.  This is accomplished by changing the total length of the inductor with a ``trombone section'', resulting in a lower inductance and therefore a higher resonant frequency.  A comb of probe signals can be sent into the device, and room temperature electronics can recover the changes in amplitude and phase without significant cross talk~\cite{Day03}, as shown in Figure~\ref{fig:detcartoon}.

\begin{figure}
\begin{center}
\includegraphics[width=1.0\columnwidth]{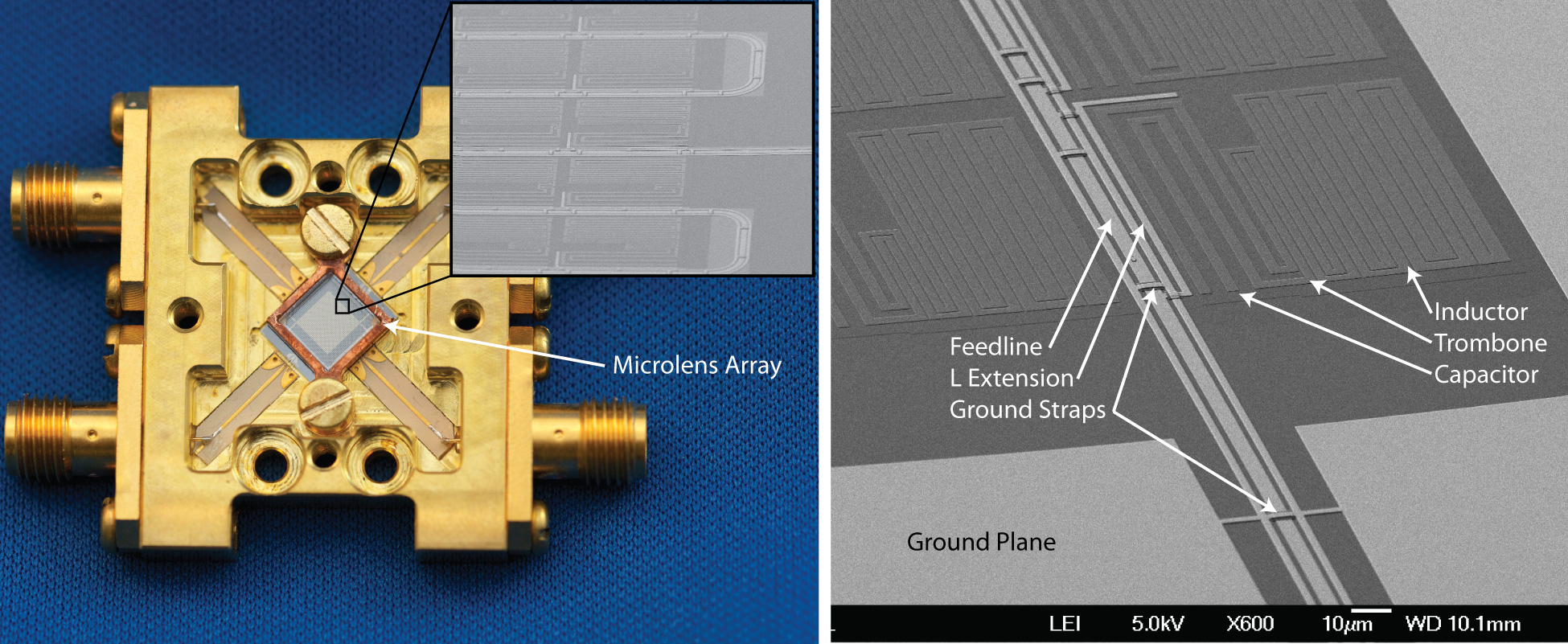}
\end{center}
\caption{Left: A photograph of the 1024 pixel OLE MKID array with microlenses mounted into a microwave package. The greyscale insets are scanning electron microscope (SEM) images of the array to show the pixel design.  The pixels are on a 100~$\mu$m pitch, with slot widths inside the resonator of 0.5~$\mu$m.  Right: A SEM of a OLE MKID pixel.  The microwave feedline runs down the middle, with ground straps shorting the finite ground planes together.  An L-shaped piece of niobium is connected to the center strip and enables strong coupling of the resonator to the feedline.  The resonant frequency is adjusted by changing the length of a ``trombone section''.   The tapering is visible as the slow increase in leg width with increasing distance from the feedline.}
\label{fig:ole}
\end{figure}

MKIDs are extremely versatile, as most resonators with a superconductor as the inductor will function as a MKID.  We have decided to pursue a lumped element resonator design~\cite{Doyle:2008p278}, shown in Figure~\ref{fig:ole}.  The resonator itself consists of a 20~nm thick sub-stoichiometric titanium nitride (TiN$_x$) film~\cite{Leduc:2010p3492}, with the nitrogen content tuned with $x<1$ such that the superconducting transition temperature $T_c$ is about 800 mK.  Due to the long penetration depth of these films ($\sim$1000 nm) the surface inductance is an extremely high 90 pH/square, allowing a very compact resonator fitting in a 100$\times$100~$\mu$m square.  Due to bandwidth limitations of our electronics we use two feedlines to read out the array, each serving 512 resonators.  The resonators are designed to be separated by 2 MHz within a 4--5 GHz band.

To avoid crosstalk between pixels the inductors are made with a double meander design that allows the electric field from the charge in each meander leg to be precisely cancelled by the adjacent leg~\cite{Noroozian:2010p4801}.  The array is designed so that resonators close together in resonant frequency are physically far apart.  To improve the quantum efficiency of the device a 100~$\mu$m pitch circular microlens array is used to focus the incoming light on the inductor, since photons hitting the capacitor or wiring will not be detected or will appear as photon events with an energy significantly below their true energy.  The circular microlenses used in these measurements limits the effective fill factor to 67\%.  An improved lens with square lens elements could increase the fill factor above 95\%.   

\begin{figure}
\includegraphics[width=1.0\columnwidth]{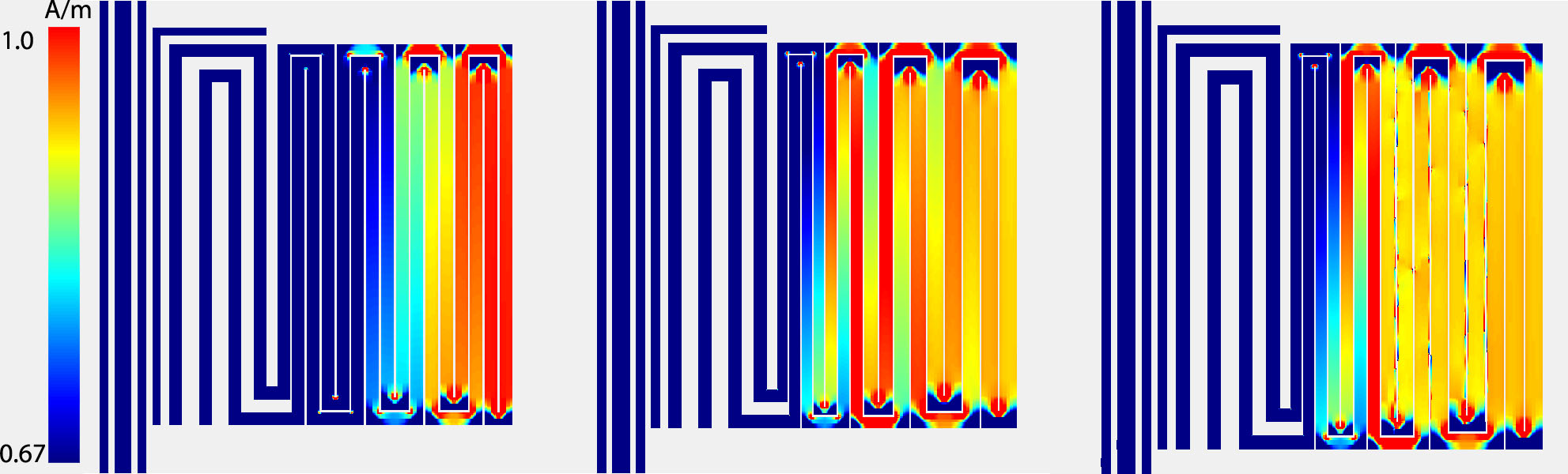}
\caption{SONNET simulations of the resonator current density.  The normalized current density is less than 0.67 in blue areas, and rises to 1.0 in the red areas.  The left panel shows a resonator with uniform leg widths.  The center panel show a resonators with rectangular legs with a mean width selected to give the most uniform current in each leg, and the right panel shows the final tapered resonator with fully optimized trapezoidal shaped legs.}
\label{fig:taper}
\end{figure}

In order to achieve high energy resolution, the OLE MKID must have a uniform response to photons hitting anywhere inside the spot produced by the microlens, which is expected to have a wavelength dependent diameter of around 15~$\mu$m.  The responsivity of an OLE MKID depends on the current density in the meander leg at the location the photon is absorbed.  Since the capacitance of our resonator is small the current density changes by nearly a factor of two over the length of the inductor.  Diffusion does not even out the quasiparticle distribution since the quasiparticle diffusion length in TiN is expected to be short (on the order of 10 $\mu$m)\cite{Leduc:2010p3492}.  In order to normalize the response we taper the width of each leg to give a uniform current density in the last eight legs (the microlens target) based on electromagnetic simulations of the current density using the SONNET software package, as shown in Figure~\ref{fig:taper}.  The widths of the legs vary from 2.5--4~$\mu$m, while the slots are 0.5~$\mu$m.

Early prototypes used coplanar waveguide (CPW) or coplanar slotline (CPS) feedlines to bring the probe signals to and from the resonators.  These designs exhibited extreme variations in coupling quality factor ($Q_c$, the strength of the resonator's coupling to the feedline) and resonant frequency, presumably due to undesired modes being excited on the feedline at discontinuities such as resonator couplers.  In order to suppress these modes we switched to a finite groundplane CPW (FCPW) feedline with regular straps that connect the ground planes together to suppress the undesired slotline mode. The feedlines are made out of Niobium since it is difficult to make 50 Ohm feedlines with TiN due to its high surface inductance.  An extended niobium groundplane was also added to reduce crosstalk between the two feedlines. In order to achieve strong coupling (low $Q_c$) a L-shaped extension of the center strip was created. 

Our OLE MKID array is fabricated on an high resistivity (10--20 kilo-ohm cm) Si $<100>$ wafer to reduce two-level system (TLS) noise~\cite{Gao:2008p66}. Wafers have the native oxide removed immediately before TiN film growth by dipping them in a buffered oxide etch (BOE). The TiN film is deposited by reactive sputtering from a 99.99\% purity Ti target in a mixture of ultra-high purity nitrogen and argon. Depositions are done at room temperature at $1\times10^{-7}$ Pa background pressure. The TiN film was deposited at a rate 37 nm/min using roughly 12\% N$_2$ in Ar by flow at 0.266 Pa total pressure. Conditions are tuned to provide a sub-stoichiometric composition with slight compressive stress ($\sim$100 MPa), Tc$\sim$800 mK, and resistivity $\sim$100~$\mu\Omega$~cm.  Layers are patterned using a Canon stepping projection aligner.  The TiN is reactive ion etched (RIE) with a chlorine containing gas mixture in an inductively coupled plasma system (ICP) using a photoresist mask. The wafer surface is solvent cleaned after etch step and given a mild O$_2$ plasma clean. TiN resonators are protected by a blanket deposit of 80 nm RF-bias sputtered SiO$_2$.  This protection layer is next patterned and etched away from transmission line areas. 

A coplanar waveguide feedline is fabricated in three steps. The center line is sputter deposited Nb which is patterned by lift-off.  An interlayer dielectric (ILD) consisting of 120 nm of SiO$_2$ is then deposited and patterned.  Finally, the surface is cleaned and patterned again for lift-off of a 160 nm Nb film used for the ground plane metal.

SiO$_2$ protecting the TiN resonators must be removed in the final step of the process for reduced TLS noise.  A stencil of positive resist is patterned over the CPW structure to protect the ILD.  The TiN is exposed by dipping the wafer in BOE for 90 seconds.  Photoresist is applied to protect the wafer for dicing into chips and not removed until the device is mounted in a sample box. 

One concern during fabrication was how accurately the tapered shape designed into the resonators would be reproduced in the final devices, since the taper is a small fraction of the line width and not visible in a standard lab optical microscope.  To measure this, devices from three separate wafer runs were measured with an scanning electron microscope (SEM).  These measurements showed that while there was an overall random error of $\sim$100 nm in the width of the slots, the widths of the legs were consistent with each other to $\sim$25 nm.  Extensive simulations show that in an ideal design with these realistic lithographic errors non-uniform current density will not limit our energy resolution until R = E/$\Delta$E exceeds 80.


\section{Experimental results}

The device was mounted in a gold-plated copper sample box and wire bonded to duroid transition boards.  The sample box was inserted into a MKID testbed based on a dilution refrigerator and cooled to approximately 100 mK.  A Weinreb microwave HEMT amplifier with a noise temperature of approximately 4 Kelvin is used to amplify the signal.  The testbed allows collimated light fed from an external fiber to illuminate the array while up to 512 resonators are probed with room temperature electronics.  The electronics consist of an Anritsu signal generators, Marki IQ mixers, and a National Instruments Analog to Digitial Converter.  More details on the electronics and cryogenics can be found in~\cite{mazinthesis}.  A mercury argon light source with narrow band filters allows illumination with monochromatic light.

\begin{figure}
\begin{center}
\includegraphics[width=0.7\columnwidth]{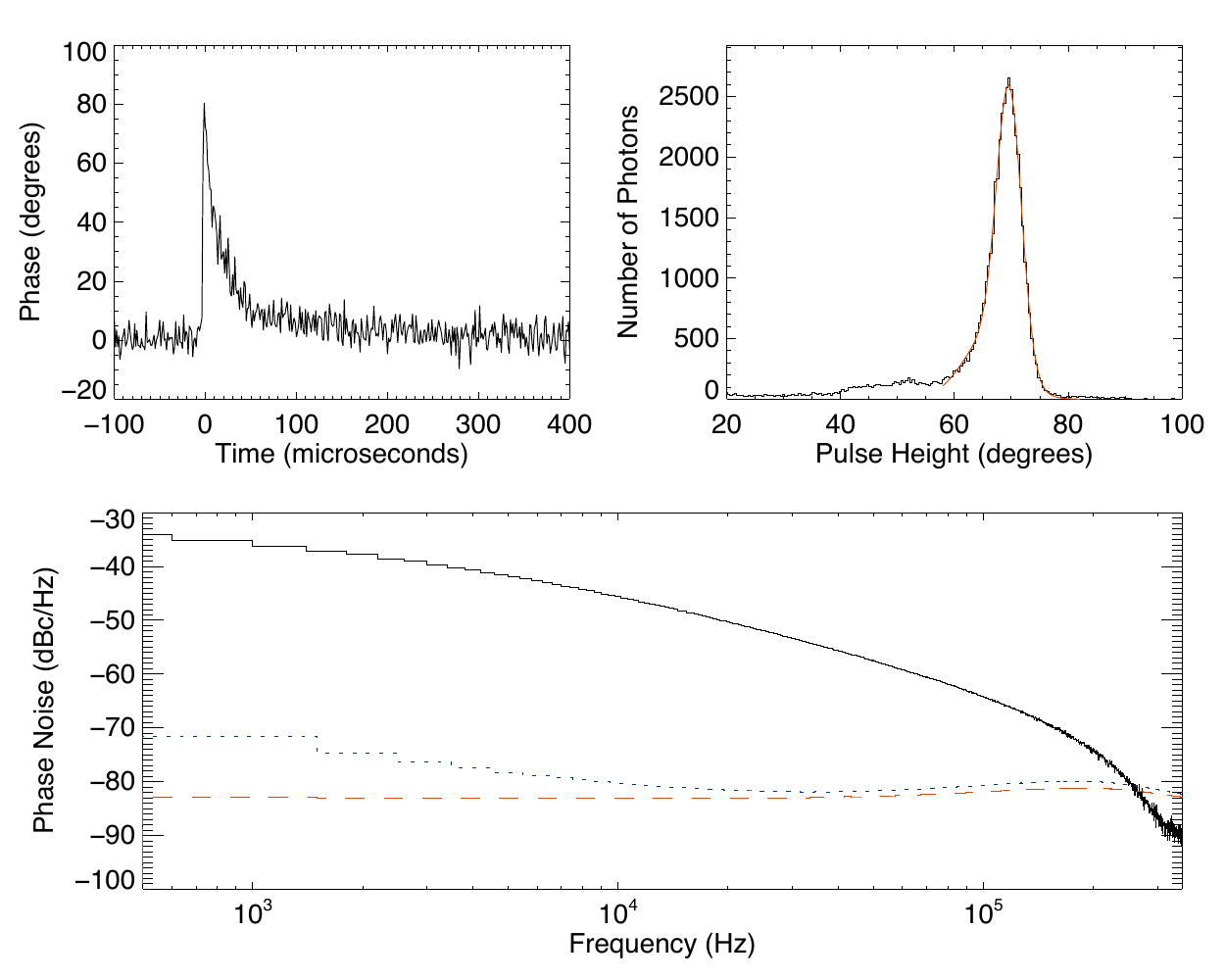}
\end{center}
\caption{The top left panel shows a characteristic single photon pulse from a 254 nm photon striking a typical resonator with measured quality factor $Q_m=18,300$.  The fall time of the pulse of 50~$\mu$s limits the maximum count to around 2000 counts/pixel/second.  The top right panel shows a histogram of the optimally estimated pulse height based on the detection of $\sim$50,000 254 nm photons.  The solid red curve is the fit of the sum of two Gaussian to the histogram, showing an energy resolution R=16, with a slightly broader shoulder extending to lower energies. The low energy shoulder is likely due to photons that miss the circular microlens and are absorbed in less sensitive areas of the resonator and photons that hit the substrate between the legs of the inductive meander.  The bottom panel shows the Fourier transform of the average pulse template in black (arbitrary scale) as well as the measured phase (blue dotted line) and amplitude (red dashed line) noise.}
\label{fig:R}
\end{figure}

The resonators were illuminated with 254 nm Hg line photons, and their response was recorded as shown in Figure~\ref{fig:R}.  After processing with a Wiener optimal filter an energy resolution R=E/$\Delta$E=16 was measured, with $\Delta$E being the FWHM.  The expected energy resolution based on the measured noise power spectrum and average pulse template was also R=16, showing good agreement with the actual measurements.  As shown in the bottom panel of Figure~\ref{fig:R}, the noise consists of both white amplifier noise and pink TLS noise exclusively in the phase direction.  Most of the signal from the photons comes at frequencies above $10^4$ Hz.  At these frequencies the phase and amplitude noise are nearly identical, indicating that the dominant noise source is the HEMT amplifier.  The white noise level of -83 dBc/Hz at the device readout power of -103 dBm is consistent with an amplifier noise temperature of 4 Kelvin assuming 3 dB of loss between the MKID and the amplifier.  Reduction of the amplifier noise temperature or an increase in the maximum usable readout power should immediately improve the energy resolution.   

In a device with a fixed response (degrees of phase shift per eV of photon energy) the energy resolution scales linearly with photon energy, except for a region between 350 and 700 nm where a significant fraction of photons pass through the metal and are absorbed in the silicon substrate.  These substrate events occur when a photon is absorbed close to the silicon/TiN interface, causing a significant fraction of the phonons ($>$70\%) created in the substrate to diffuse into the TiN and break Cooper Pairs.  These substrate events can give quite large signals, and will be discussed in depth in a future paper.  Simple device changes, such as making the TiN film thicker or using a transparent substrate such as sapphire, should eliminate these unwanted substrate events.

Significant improvements can be made to increase the energy resolution, as the theoretical energy resolution set by the creation statistics of the quasiparticles created during downconversion is $R = \frac{1}{2.355} \sqrt{\frac{ \eta h \nu}{F \Delta }}$, where $\eta=0.57$ is the efficiency of creating quasiparticles~\cite{Kozorezov:2007p763}, $h \nu$ is the energy of the
incident photon, $\Delta=1.72 k_B T_c$ is the gap energy of the superconducting
absorber, and $F \approx 0.2$ is the Fano factor~\cite{fano47}.  This works out to R=150 at 5 eV for an operating temperature of 100 mK.  An operating temperature of 15 mK could allow a theoretical maximum energy resolution of R=400 at 5 eV, although it is likely other noise sources, like two level system noise~\cite{Gao:2008p66}, will become more important as future development increases the energy resolution.    

\begin{figure}
\begin{center}
\includegraphics[width=0.7\columnwidth]{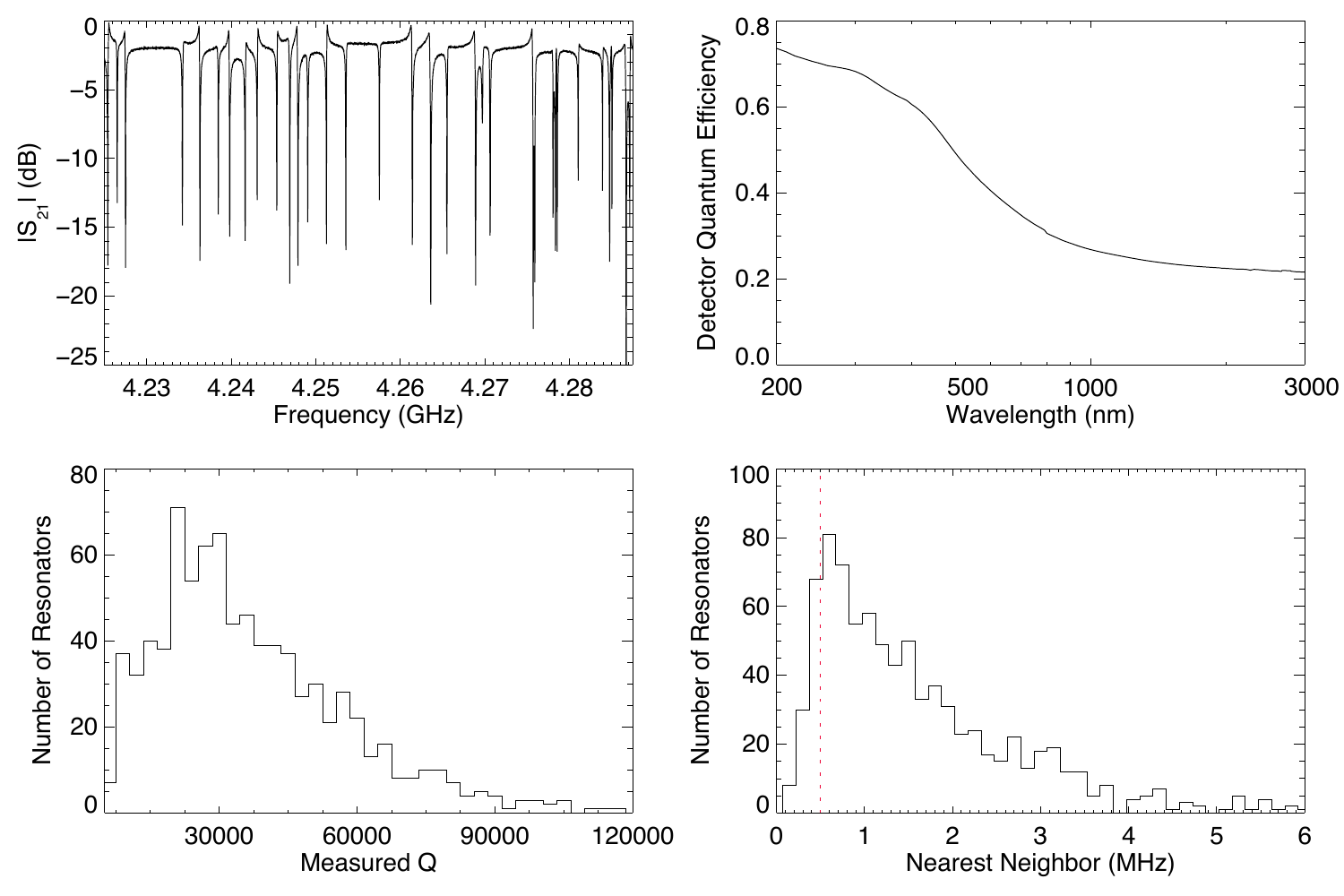}
\end{center}
\caption{The top left panel shows the microwave transmission through the device over 10\% of the frequency span covered by resonators.  The top right panel is a measurement of the quantum efficiency of a bare 40 nm TiN on sapphire film.  The bottom left panel is a histogram of measured quality factor for 852 out of a possible 1024 resonators.  The internal quality factor of the resonators, $1/Q_i = 1/Q_m - 1/Q_c$, was approximately $1\times10^6$.  The bottom right panel is the frequency spacing in MHz between each resonator and its nearest neighbor.  Most of the missing resonators are too close together in frequency ($<500$~kHz, noted with a dashed line), resulting in only one resonator being included in the plot. }
\label{fig:meas}
\end{figure}

Figure~\ref{fig:meas} shows data from a typical 32x32 pixel device.  In this device, 85\% of the resonators were usable.  Simple number counts showed that at least 95\% of the resonators were present, but variations in thickness and T$_c$ of the TiN film caused some resonators to have similar resonant frequencies and overlap in frequency.  Future improvement to the uniformity of the TiN and a more robust algorithm for placement of resonators should significantly decrease the number of overlapping resonators.

The quantum efficiency of the device was estimated by depositing a 40 nm TiN film on sapphire and measuring the reflection and transmission of the film as a function of wavelength, then subtracting these quantities from unity to give the fraction of photons absorbed in the superconducting film, as shown in the top right panel of Figure~\ref{fig:meas}.  This was done with a Varian Cary 5000 spectrometer with accessories for absolute transmission and absolute specular reflectance.  This figure slightly overestimates the quantum efficiency of the final device because it does not account for losses in the microlens array, or the $\sim$10\% correction due to photons below 0.5~$\mu$m slipping through the slots in the inductor.  The quantum efficiency is extremely good in the UV, but declines to about 30\% at 1~$\mu$m.  Further development will likely result in significant increases in quantum efficiency.   



\section{Conclusion}

Improvements in the fabrication of TiN films will lead to increased uniformity and fabrication yield.  The array size can be grown nearly arbitrarily by adding more resonators per feedline, likely to a maximum of around 10,000 resonators in a 10--20 GHz band, and then by adding more microwave feedlines.  The primary challenges to growing the array are in the digitization bandwidth and processing power of the room temperature electronics, not the device fabrication.  

OLE MKID arrays have now been proven in the lab, and the first astronomical test at the Palomar 200 inch telescope has been conducted with data taken on a variety of astronomical objects.  These arrays will bring extreme performance improvements to some of the most exciting areas of astrophysics, such as coronagraphic planet finding~\cite{Crepp:2011p4649}, transient and time variable sources~\cite{Romani:2001p1716}, and high redshift galaxy evolution~\cite{Bouwens:2008p2309}, and will likely also find application in other fields such as quantum optics~\cite{Ma:2009p4760} and biological imaging~\cite{Tinoco:2011p4660}.

This material is based upon work supported by the National Aeronautics and Space Administration under Grant NNX09AD54G, issued through the Science Mission Directorate, Jet Propulsion Lab's Research \& Technology Development Program, and a grant from the W.M. Keck Institute for Space Studies.  Part of the research was carried out at the Jet Propulsion Laboratory, California Institute of Technology, under a contract with the National Aeronautics and Space Administration.  The authors would like to thank Rick LeDuc, Jonas Zmuidzinas, Sunil Golwala, David Moore, Peter Day, and Omid Noroozian for useful insights.

\end{document}